\documentclass[3p,times,procedia,margin=1in]{elsarticle}
\usepackage{ecrc}
\usepackage[bookmarks=false]{hyperref}
    \hypersetup{colorlinks,
      linkcolor=blue,
      citecolor=blue,
      urlcolor=blue}

\volume{00}

\firstpage{1}

\journalname{Procedia Computer Science}

\runauth{Author name}

\jid{procs}

\usepackage{amssymb}
\usepackage{amsmath}
\usepackage{booktabs}
\usepackage{multirow} 
\usepackage{graphicx}
\usepackage{caption}
\usepackage{tabularx}
\usepackage{subcaption}
\usepackage{geometry}
\usepackage{makecell}
\bibliography{refs}  
\usepackage{setspace}

\usepackage[figuresright]{rotating}

\begin{document}
\begin{frontmatter}



\dochead{Proceedings of the 2025 International Conference on Biomimetic Intelligence and Robotics}

\title{Unlocking Mixed Reality for Medical Education:\\A See-Through Perspective on Head Anatomy}


\author[a]{Yuqing Wei\fnref{equal}} 
\author[a]{Yupeng Wang\fnref{equal}} 
\author[a]{Jiayi Zhao}
\author[a]{Yanjun Liu}
\author[a]{Huxin Gao}
\author[a]{Jiewen Lai\corref{cor2}}

\address[a]{Department of Electronic Engineering, The Chinese University of Hong Kong, Hong Kong SAR, China}

\begin{abstract}
Extended reality (XR), encompassing Virtual Reality (VR), Augmented Reality (AR), and Mixed Reality (MR), is emerging as a transformative platform for medical education. Traditional methods such as textbooks, physical models, and cadaveric dissections often lack interactivity and fail to convey complex spatial relationships effectively. The emerging MR technology addresses these limitations by providing immersive environments that blend virtual elements with real-world contexts. This study presents an MR application for head anatomy education, enabling learners to intuitively interact with see-through 3D anatomical structures via hand gestures and controllers. Our hierarchical information design supports progressive learning, guiding users from basic anatomical labels to detailed structural insights. Additionally, the system incorporates an automatic calibration module that aligns virtual anatomical models with a real human head, thereby facilitating realistic human-model interactions. Experiments show that the system can effectively match the anatomical model with real-time scenes, thus enhancing the interactivity and immersion of medical education, providing an innovative tool for teaching anatomy.
\end{abstract}


\cortext[cor2]{Corresponding author. Email: jwlai@ee.cuhk.edu.hk}
\fntext[equal]{Equal contribution.}

\end{frontmatter}


\vspace*{-15pt}

\section{Introduction}
\vspace{-10pt}
\label{main}

With the rapid development of computer technology, extended reality (XR), such as virtual reality (VR), augmented reality (AR), and mixed reality (MR), is increasingly being used in education and medicine. VR creates a fully synthetic interactive environment that immerses users in a computer-generated space \cite{b1}. In contrast, AR overlays digital elements onto the physical world, allowing users to perceive and interact with both real and virtual objects \cite{b2}. MR combines the capabilities of VR and AR, merging the physical and digital environments to enable real-time interaction with 3D virtual content in the real world \cite{b4}. Over the past few years, MR has garnered increasing attention in the medical field \cite{b5,b6}, as visualization and spatial perception of complex anatomical structures are crucial. MR is a hybrid of AR and VR, resulting from the integration of the physical and digital worlds \cite{b7,b8}. In this case, it addresses the limitations of VR, which excludes the real-world environment, and AR, which cannot interact with three-dimensional (3D) data packages. In the medical field, MR has great potential in enriching the real-world environment with virtual data on a single display and allows interaction through various methods \cite{b9}.

Traditional medical education relies on textbooks, physical models, and cadaveric specimens. However, these approaches present several limitations, including low interactivity, constrained contextual environments, and challenges in visualizing complex anatomical structures underneath. Recent studies indicate that both students and educators perceive MR-assisted instruction as more effective than conventional methods in medical teaching \cite{b10}. MR enables a more intuitive and immersive learning experience by allowing students to interact with virtual 3D anatomical content in real time and observe internal structures layer by layer to gain a better understanding of spatial relationships that are often difficult to grasp through traditional static resources.

\begin{figure}[t]
    \centering
    \includegraphics[width=\linewidth]{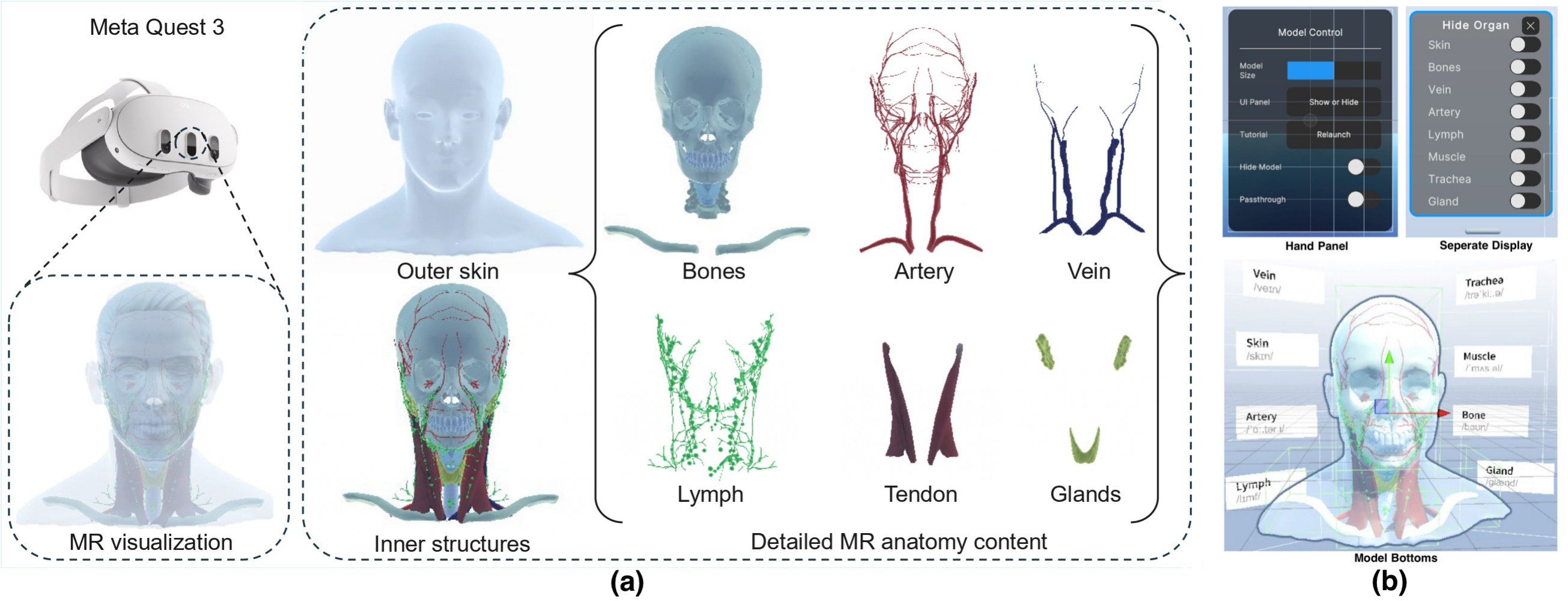}
    \vspace{-20pt}
    \caption{(a) Overview of MR anatomy education. (b) User Interface.}
    \label{fig:Overview}
    \vspace{-20pt}
\end{figure}

Based on these advantages, this study introduces an interactive MR application designed to enhance immersive teaching and intuitive learning of head anatomy through a see-through perspective. The application enables users to visualize detailed 3D anatomical structures superimposed on the upper chest area, while simultaneously maintaining awareness of their physical environment via a head-mounted display (Meta Quest 3). A schematic diagram, as shown in Fig.\ref{fig:Overview}, outlines this work.

To enable accurate spatial alignment, an estimation of the head pose based on MediaPipe \cite{b11} is incorporated into an automatic calibration module. This module aligns the virtual anatomical model with a real-world anatomical reference, providing a more intuitive and context-aware learning experience. Additionally, users can interactively explore anatomical features by adjusting the model’s transparency, scale, and the visibility of specific organ regions. Textual descriptions are integrated to supplement each region, offering layered anatomical knowledge to support self-paced learning. Our approach provides a valuable reference for immersive medical education using MR technology, with the potential to be extended to other anatomical regions and organ systems.

The main contributions of this paper are as follows:
\vspace{-10pt}
\begin{itemize}
    \item We propose a MR-based medical education platform focused on head anatomy. The system offers an interactive learning interface that supports manipulation of 3D models and multilayered anatomical exploration, thereby promoting spatial understanding and learner engagement.
    
    \item We design and implement an automatic calibration module that spatially aligns a virtual anatomical model with a real human head. This feature supports personalized anatomical education directly on real individuals, offering promising potential to enhance educational value.
    
    \item We conducted an experimental validation with three participants to assess the system’s usability and accuracy. Both qualitative and quantitative results suggest the system holds promise for enhancing medical education.
\end{itemize}

\vspace{-25pt}
\section{Related Work}
\vspace{-10pt}
VR immerses users in a fully synthetic environment through head-mounted displays (HMDs), enabling complete detachment from the physical surroundings. This technology has found extensive applications in education, healthcare, engineering, and other domains. Coyne \textit{et al.} suggests that VR technology is an immersive, comfortable, and engaging learning platform that can be used for team-based remote learning \cite{b12}. VR not only enables experiential learning by providing simulated virtual environments, but also supports the creation of works in these environments and triggers activities through user interaction, supporting dynamic forms of learning \cite{b13}. For instance, product design can leverage VR technology to create complex 3D models and products using 3D modeling tools and visualize them, facilitating interactive and real-time communication of product information \cite{b14}. VR is also applied in the medical field, particularly in surgery and education. In surgical training, VR technology can significantly reduce the likelihood of surgical errors, improving patient safety \cite{b15}. The use of VR technology also aids rehabilitation clinicians in remote therapy---telerehabilitation, allowing patients to exercise in a virtual environment at home and transmit data to their doctors \cite{b16}. In medical education, Haluck and Krummel describe a progressive learning model in which VR-based simulation learning is a crucial step. Students can practice components of a procedure or the entire operation before conducting the final stages of skills training in the operating room \cite{b17}, which helps them gain confidence before the resident rotations. 

In contrast, AR is an interactive display environment based on the real world, enhancing the user's experience by utilizing computer-generated displays, text, sound, and special effects. AR supports the enhancement of the user's view by overlaying data and information in real-time in the actual environment \cite{b18}. AR has been widely applied in the medical field. In medical rehabilitation, Coelho \textit{et al.} created a physical mixed model and a mobile device AR application using imaging data from a patient who had undergone surgery for cranial deformity correction. This method was used by 38 experienced surgeons who considered it a valuable educational tool for teaching and pre-surgical planning \cite{b19}. Ferrari \textit{et al.} utilized electromagnetic sensors to detect nickel-titanium tubes---one of the commonly used medical devices, enabling AR visualization of hidden deformable tubular structures \textit{in situ}. The obtained alignment accuracy demonstrated the feasibility of this method, which can be used in advanced AR simulations, particularly as a tool for identifying and isolating tubular structures \cite{b20}. Ortiz-Catalan \textit{et al.} employed AR technology to enable amputees to see a virtual arm appear on a screen, which could be controlled by the brain to move the previously severed limb, thus achieving therapeutic results \cite{b21}.

MR combines elements of both VR and AR, containing both physical and virtual components \cite{b22}. MR is often described by experts as a sliding scale between an entirely physical environment with no virtual elements and a fully virtual environment \cite{b9}. MR is not confined to the virtual environments of VR or the single mode of interaction found in AR. Parveaua and Addaa categorized MR into three types: First, it consists of both real and virtual content, allowing data contextualization. Second, the digital content requires real-time interaction. Third, the content must be spatially mapped and related to three-dimensional space \cite{b23}. MR technology is promising for exploring the anatomical structure of the human body. McJunkin \textit{et al.} demonstrated that MR headsets can display interactive 3D anatomical models of the temporal bone, anchored in real space. By manually aligning the fixed model with a real head, this technology shows potential as an image-guided tool for anatomy and lateral skull base surgeries \cite{b25}. In the field of tumor surgery, Scherl \textit{et al.} applied MR in preoperative planning for minimally invasive electroporation and microwave ablation of advanced hepatic gastrointestinal tumors, optimizing surgical methods through remote and hospital-based analysis of the patient's specific anatomical structures \cite{b26}. In otolaryngology and maxillofacial surgery, MR facilitates easier extrafascial dissection of parotid tumors, thereby reducing the incidence of complications \cite{b27}. MR-guided liver resection planning has improved understanding of liver vascular anatomy and tumour location, enhancing resection accuracy while preserving a larger residual liver volume \cite{b28}. The development of patient-specific VR and MR models using web-based services and their preoperative and intraoperative application demonstrates the significant potential of the surgical assistance tools \cite{b29}.

\begin{figure*}[t]
    \centering
    \includegraphics[width=0.83\linewidth]{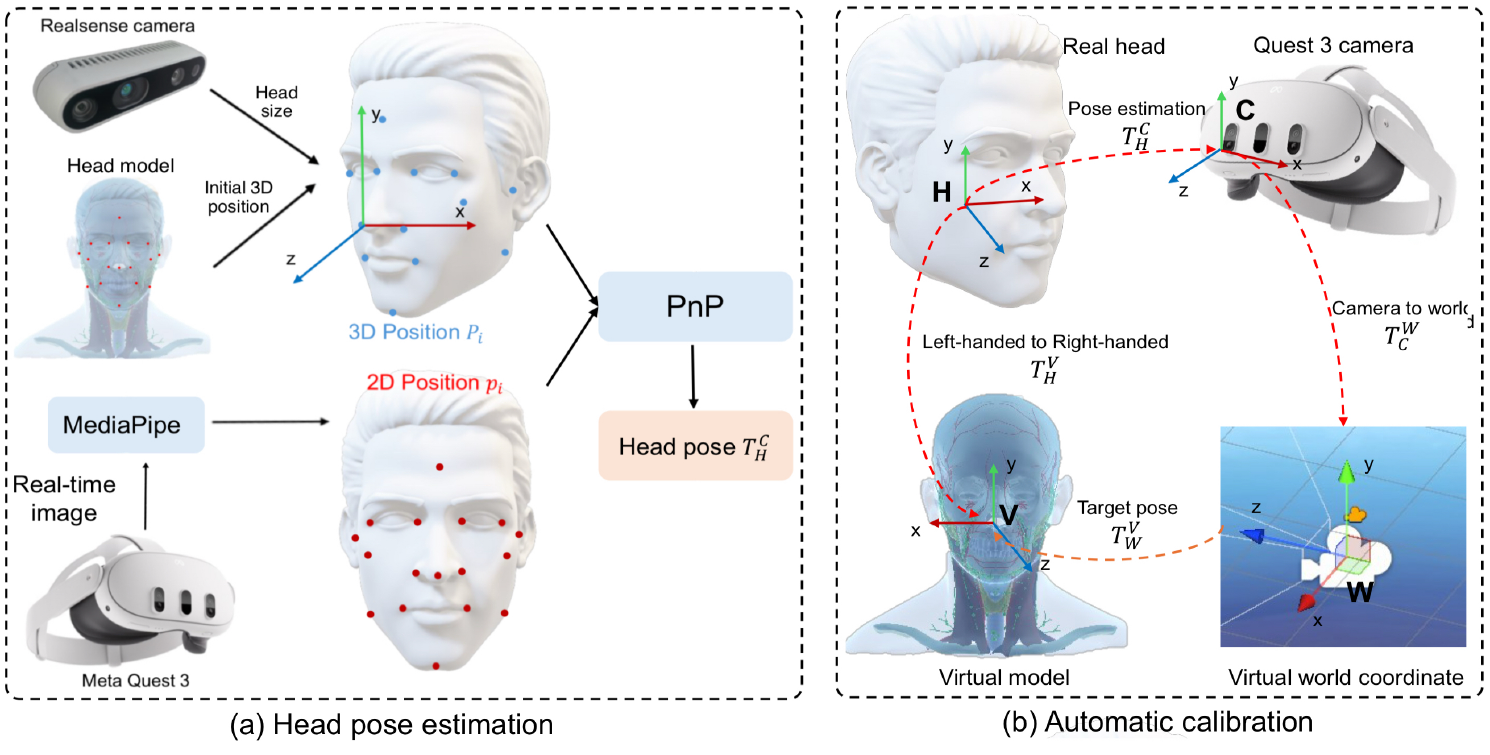}
    \vspace{-10pt}
    \caption{Automatic virtual-to-real calibration. (a) The head pose estimation procedure. (b) The kinematic chain of the calibration system. Red dashed lines indicate real-time transformations; the orange dashed line denotes the target pose to compute.}
    \label{fig:Calibration}
    \vspace{-30pt}
\end{figure*}
\vspace{-10pt}

\section{Methodology}
\vspace{-10pt}
\subsection{ Overview}
\vspace{-10pt}
The proposed application is deployed on the Meta Quest 3 platform by using Unity 3D, enabling users to interact with an accurate 3D model of the human head in an immersive MR environment. As illustrated in Fig.\ref{fig:Overview}, the system presents a detailed anatomical structure that includes multiple organ systems and tissue types such as skin, bones, arteries, veins, lymphatic vessels, glands, and tendons. The 3D anatomical models used in this study were sourced from the project ``Lymphatics Drainage of Head and Neck'', developed by the University of Dundee, School of Dentistry \cite{b30}, and are licensed under CC BY-SA 4.0.

To enhance anatomical clarity and usability, the application adopts a color-coded visualization scheme. This visual differentiation aids users in quickly identifying specific structures and understanding their spatial relationships. The model supports dynamic visualization, allowing learners to view external features like skin as well as internal organs layer by layer. The system includes several core features aimed at promoting interactive learning:

\textbf{(1) Gesture and Controller-Based Manipulation:} Users can zoom, rotate, drag, and place the 3D anatomical model using either hand gestures or handheld controllers for flexible interaction.

\textbf{(2) Detailed Anatomy Explanation:} Each part of the head can be individually visualized, allowing users to isolate specific structures. For each part, corresponding educational content is provided, including detailed descriptions, functions, and related anatomical relationships.

\textbf{(3) Real-World Model Alignment:} An automatic calibration module enables spatial alignment of the virtual anatomy with a physical anatomical reference. This helps users relate digital content with real-world context and enhances comprehension of internal anatomical structures.

With the above features, this MR-based application offers an intuitive and engaging learning tool to support medical education.

\vspace{-10pt}
\subsection{ Automatic Calibration}
\vspace{-10pt}
In addition to directly visualizing the virtual head model, the system is capable of automatically aligning the model with a real human head. 
\vspace{-10pt}

\subsubsection{Head Pose Estimation}
Initially, as illustrated in Fig.\ref{fig:Calibration} (a), the head pose is estimated using a combination of MediaPipe for facial landmark detection and the Perspective-n-Point (PnP) algorithm for 3D pose estimation. To apply the PnP algorithm, a set of 2D–3D point correspondences ($p_j \leftrightarrow P_j$) is required. 

To obtain the 3D points $P_j$, the initial 3D coordinates $P_o$ of key facial landmarks are defined on a virtual anatomical model with dimensions of $16\times 23~cm$. The coordinate system is established with the nose tip as the origin, and all axes are aligned with the facial orientation. Next, a scaling factor derived a depth camera will be used to adjust $P_o$ to better approximate the real facial 3D coordinates. 

In this study, an RGB-D camera (Intel RealSense D435) is used to capture both the RGB image $I$ and the aligned depth image $D$. The RGB image is processed by MediaPipe, which outputs a set of 2D facial landmark positions $\{p_k = (u_k, v_k)\}$. For each 2D point $p_k$, the corresponding 3D point $P_k$ in the camera coordinate system is reconstructed using the depth image and the camera intrinsic matrix $K$ as follows:
\vspace{-20pt}
\begin{equation}
    P_k = D(u_k, v_k) \cdot K^{-1} 
    \begin{bmatrix}
        u_k \\
        v_k \\
        1
    \end{bmatrix},
    \vspace{-20pt}
\end{equation}
where $D(u_k, v_k)$ denotes the depth value at pixel $(u_k, v_k)$, and $K$ is the camera intrinsic matrix defined as:
\vspace{-20pt}
\begin{equation}
    K = 
    \begin{bmatrix}
        f_x & 0 & c_x \\
        0 & f_y & c_y \\
        0 & 0 & 1
    \end{bmatrix}.
    \vspace{-20pt}
\end{equation}
Here, $(f_x, f_y)$ are the focal lengths and $(c_x, c_y)$ represents the principal point of the camera. 

The width ($w$) and length ($l$) of the face can be easily obtained with a few reconstructed 3D points $P_k$. Then, the scaling matrix $A$ is constructed as follows:
\vspace{-20pt}
\begin{equation}
\label{eq:scaling_matrix}
\begin{aligned}
    A &= 
    \begin{bmatrix}
        A_x & 0 & 0\\
        0 & A_y & 0\\
        0 & 0 & A_z\\
    \end{bmatrix}, \quad 
    A_x = \frac{w}{w_o}, \quad A_y = \frac{l}{l_o}, \quad A_z = \frac{A_x + A_y}{2}.
\end{aligned}
\vspace{-20pt}
\end{equation}
where $w_o$ and $l_o$ are the width and length of the anatomical model. Therefore, the 3D points of the real face can be approximated as:
\vspace{-20pt}
\begin{equation}
    P_j = 
    P_o\cdot A
    \vspace{-20pt}. 
\end{equation}

Since the target head remains unchanged, when observed by the Meta Quest 3, the 3D position satisfies $P_i = P_j$. The image captured by the Meta Quest 3 is also processed by MediaPipe, yielding the 2D positions of facial landmarks $\{p_i\}$. Using the previously obtained 3D points $\{P_i\}$ and their corresponding 2D image projections $\{p_i\}$, the PnP algorithm is employed to estimate the rotation and translation vectors that define the head pose $T_H^C$ with respect to the Meta Quest 3 camera.
\vspace{-10pt}

\subsubsection{Virtual-to-real Calibration}
With the head pose and head size, the anatomical model can be adjusted to fit the real face. The scaling factor obtained from the depth camera is again used to adjust the model size. To maintain its shape, the anatomical model is uniformly scaled along three axes, with a factor of $A_z$ in Eq.~\ref{eq:scaling_matrix}. 

The pose of the virtual model $V$ in the virtual world frame $W$, denoted as $T_W^V$, must be computed to accurately overlay the virtual model onto the real head. As illustrated in Fig.~\ref{fig:Calibration}(b), the corresponding kinematic chain is established. The system contains two types of coordinate frames: the virtual components developed in Unity use a left-handed coordinate system, while the real components are defined in a right-handed coordinate system.

Therefore, the first step is to convert the left-handed frame to a right-handed one. Specifically, this is achieved by inverting the $z$-axis of the virtual model frame $V$. Afterwards, a rotation of $180^\circ$ about the $y$-axis is applied, which maps the left-handed virtual model frame $V$ to the right-handed frame $H$, ensuring consistency between the coordinate systems of the virtual and real environments.

In addition, since the Unity application is deployed directly on the Meta Quest 3 device, it provides a direct transformation from the camera frame $C$ to the virtual world frame $W$ by utilizing the onboard Simultaneous Localization and Mapping (SLAM) system. As a result, the transformation $T_W^C$ can be obtained in real time. Finally, with the estimated pose of the head $H$ in the camera frame $C$, the target pose of the virtual model in the world frame can be calculated as:
\vspace{-20pt}
\begin{equation}
    T_W^V = T_H^V \left( T_H^C T_C^W \right)^{-1}.
    \vspace{-20pt}
\end{equation}

This equation enables precise overlay and registration of the virtual model onto the real head within the MR environment.
\vspace{-10pt}

\subsection{ Education Features}
\vspace{-10pt}
To enhance user engagement and facilitate anatomical learning, the system integrates a set of educational features designed specifically for MR environments. As shown in Fig.\ref{fig:Overview}(b), these features include interactive information display, selective visualization of anatomical regions, and an intuitive user interface (UI) for control and guidance.

The application supports selective visualization, allowing users to toggle the visibility of individual tissues through a control panel. Transparency adjustments are also available, enabling inspection of deeper anatomical layers beneath surface structures. An interactive wrist panel is mounted on the user’s virtual wrist, providing quick access to key functionalities. Upon launch, users are guided by an in-scene tutorial video and an optional onboarding panel, both of which can be reactivated as needed.

These integrated educational features create an immersive and learner-centered MR experience that supports both exploratory learning and structured knowledge acquisition.

\vspace{-10pt}
\section{Experiments and Results}
\vspace{-10pt}
One of the significant features of the proposed system is its ability to anchor anatomical content directly onto a real human head. This personalized visualization enhances educational value by allowing users to observe anatomical structures overlaid on actual individuals in real-time. To evaluate this function, three participants were recruited, each tested under various head poses. Both qualitative and quantitative analyses were conducted to assess the alignment accuracy and robustness of the overlaid anatomical content. 

\vspace{-10pt}
\subsection{ Qualitative Evaluation}
\begin{figure*}[t]
    \centering
    \includegraphics[width=\linewidth]{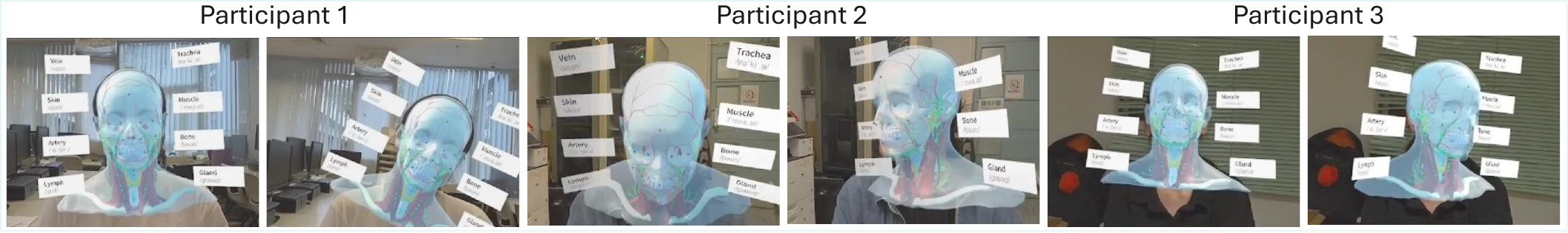}
    \vspace{-20pt}
    \caption{Qualitative evaluation: overlay of the virtual anatomical model onto real heads under various poses.}
    \label{fig:Results}
    \vspace{-15pt}
\end{figure*}

\vspace{-10pt}
As shown in Fig.~\ref{fig:Results}, a user wearing the MR device was instructed to observe the heads of three participants. The system then automatically aligned the 3D anatomical model to the detected head pose in real time. From the captured frames, it can be seen that the virtual anatomical structures align reasonably well with the real-world head positions from multiple viewpoints. Furthermore, the overlay remained stable under moderate head rotations and translations, suggesting the robustness and reliability of the virtual-to-real calibration method used in the system.

\vspace{-10pt}
\subsection{ Quantitative Evaluation} 
\begin{figure}[h]
\centering
\begin{minipage}{0.38\textwidth}
    \centering
    \includegraphics[width=0.5\linewidth]{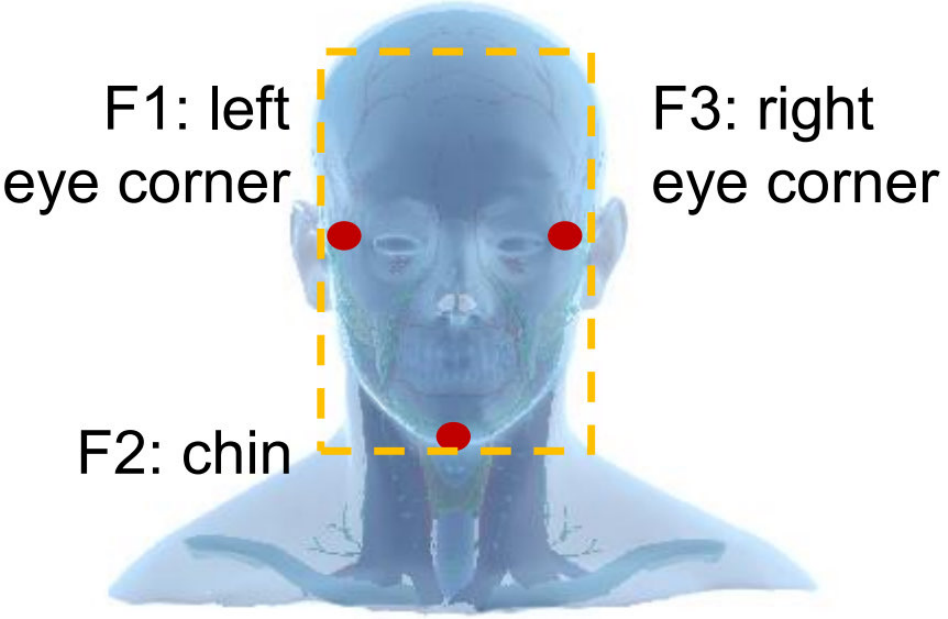}  
    \vspace{-10pt}
    \caption{Key facial landmarks.}
    \label{fig:key_facial_landmarks}
\end{minipage}
\begin{minipage}{0.48\textwidth}
    \centering
    \renewcommand{\arraystretch}{1}  
    \setlength{\tabcolsep}{15pt} 
    \scriptsize 
    \captionof{table}{Head sizes of participants.}
    \vspace{-5pt}
    \begin{tabular}{l|c|c}
        \toprule
        \textbf{Participants} & \textbf{Head Size ({\it{px}})} & \textbf{Head Size ({\it{mm}})}\\
        \colrule
        \textbf{Participant 1 (P1)} & 653.32 $\times$ 737.69 &  125 $\times$ 145\\
        \textbf{Participant 2 (P2)} & 793.66 $\times$ 886.12 &  125 $\times$ 150\\
        \textbf{Participant 3 (P3)} & 596.75 $\times$ 678.02 &  110 $\times$ 133\\
        \bottomrule
    \end{tabular}
    \label{tab:headsize}
\end{minipage}
\vspace{-25pt}
\end{figure}

\vspace{-10pt}
To quantitatively assess the alignment precision, we evaluated overlay errors at three key facial landmarks: the left eye corner (F1), chin (F2), and right eye corner (F3), as illustrated in Fig.~\ref{fig:key_facial_landmarks}. Table~\ref{tab:headsize} demonstrates the estimated head size in image space and the corresponding real head size of different participants.

A total of 166 head poses from three participants were used for analysis. For each pose, the landmarks F1, F2, and F3 were manually annotated on the overlaid virtual model in recorded videos. Simultaneously, the same video frames without the overlay were processed using the MediaPipe framework to automatically detect corresponding facial landmarks. The overlaying error was then calculated as the Euclidean distance between the positions of each manually annotated landmark on the virtual model and the automatically detected landmark on the real face.

The results are summarized in Table~\ref{tab:overlay_error}, which shows the mean pixel errors of the three key landmarks on different participants. Based on the average pixel-to-millimeter ratio from the head dimensions, the corresponding errors in millimeter are estimated. Over the three participants, the mean errors were 7.73$\pm$4.36~mm for F1 (left eye corner), 5.69$\pm$3.29~mm for F2 (chin), and 5.79$\pm$3.42~mm for F3 (right eye corner). Therefore, the overall error of the three landmarks was 6.40$\pm$3.81~mm in physical space. Compared with the AR visualization error of 6.97~mm from \cite{b31} and the misalignment error of 16.50~mm from the method based on reflective-AR display \cite{b32}, the findings indicate that the system achieves a reasonable level of anatomical alignment suitable for educational purposes in real-world conditions.
\vspace{-20pt}

\begin{table}[ht]
\centering
\renewcommand{\arraystretch}{0.9}  
\setlength{\tabcolsep}{0.35pt}       
\scriptsize                   
\caption{Overlaying error for different facial features.}
\vspace{-5pt}
\begin{tabular}{l c | cccc | cccc | cccc | cccc}
\toprule
 & \makecell[c]{\textbf{Number of}\\ \textbf{head poses}}
 & \multicolumn{4}{c|}{\textbf{F1}} 
 & \multicolumn{4}{c|}{\textbf{F2}} 
 & \multicolumn{4}{c|}{\textbf{F3}} 
 & \multicolumn{4}{c}{\textbf{All}} \\
\cmidrule(lr){3-6} \cmidrule(lr){7-10} \cmidrule(lr){11-14} \cmidrule(lr){15-18}
 & & mean(px) & mean(mm) & std(px) & std(mm)
   & mean(px) & mean(mm) & std(px) & std(mm)
   & mean(px) & mean(mm) & std(px) & std(mm)
   & mean(px) & mean(mm) & std(px) & std(mm) \\
\midrule
\textbf{P1} & 71  & 33.74 & 6.46 & 15.92 & 3.05 & 25.53 & 4.89 & 11.98 & 2.29 & 26.57 & 5.08 & 13.32 & 2.55 & 28.61 & 5.47 & 14.22 & 2.72 \\
\textbf{P2} & 39  & 62.16 & 9.79 & 23.72 & 3.74 & 41.64 & 6.56 & 18.79 & 2.96 & 40.35 & 6.36 & 20.70 & 3.26 & 48.05 & 7.57 & 23.16 & 3.65 \\
\textbf{P3} & 56  & 42.91 & 7.91 & 29.70 & 5.47 & 33.13 & 6.11 & 22.94 & 4.23 & 34.12 & 6.29 & 22.94 & 4.23 & 36.72 & 6.77 & 25.55 & 4.71 \\
\textbf{Mean} & & & 7.73 & & 4.36 & & 5.69 & & 3.29 & & 5.79 & & 3.42 & & \textbf{6.40} & & \textbf{3.81} \\
\bottomrule
\end{tabular}
\label{tab:overlay_error}
\vspace{-15pt}
\end{table}

\vspace{-20pt}
\section{Discussion and Conclusion}
\vspace{-10pt}
This study presents the development of an MR application aimed at enhancing the understanding of head anatomy. By enabling real-time interaction with 3D anatomical models, the system provides an immersive and intuitive learning and teaching experience. A key innovation of the proposed system is the integration of an automatic calibration module that spatially aligns virtual anatomical structures with a real human head, enhancing contextual relevance and educational value.

However, the system currently lacks the accurate 3D coordinates of real facial landmarks, which may cause slight alignment deviations. Additionally, the reliance on a pre-defined 3D head model may exhibit significant misalignment when applied to individuals with different facial shapes. A potential solution to this problem is to adjust the three dimensions of the model with independent scaling factors, enhancing the model's flexibility to some extent. Future work will focus on integrating high-precision depth-sensing technologies to improve the accuracy of landmark localization, as well as including a more diverse participant pool in terms of skin tone and facial shape to improve its generalizability. Furthermore, participants' knowledge levels will be assessed before and after using the application to evaluate its educational effectiveness.
\vspace{-30pt}
\section*{Acknowledgements}
\vspace{-10pt}
This work is partially supported by the Guangdong Basic and Applied Basic Research Foundation General Program under grant 2025A1515011594; in part by the Hong Kong Research Grants Council Collaborative Research Fund under grant C4026-21GF; in part by CUHK--Department of Electronic Engineering MSc Research \& Development Project grant.




\vspace{-11pt}
 \begin{thebibliography}{}
\vspace{-10pt}
\footnotesize
\setstretch{0.95}

\bibitem{b1} W. S. Khor, B. Baker, K. Amin, A. Chan, K. Patel, and J. Wong, ‘Augmented and virtual reality in surgery—the digital surgical environment: applications, limitations and legal pitfalls’, \textit{Ann. Transl. Med.}, vol. 4, no. 23, pp. 454–454, Dec. 2016.
\bibitem{b2} R. T. Azuma, ‘A Survey of Augmented Reality’, \textit{Presence Teleoperators Virtual Environ.}, vol. 6, no. 4, pp. 355–385, Aug. 1997.
\bibitem{b4} Å. Fast-Berglund, L. Gong, and D. Li, ‘Testing and validating Extended Reality (xR) technologies in manufacturing’, \textit{Procedia Manuf.}, vol. 25, pp. 31–38, 2018.
\bibitem{b5} J. Han, H.-J. Kang, M. Kim, and G. H. Kwon, ‘Mapping the intellectual structure of research on surgery with mixed reality: Bibliometric network analysis (2000–2019)’, \textit{J. Biomed. Inform.}, vol. 109, p. 103516, Sep. 2020.
\bibitem{b6} A. W. K. Yeung \textit{et al.}, ‘Virtual and Augmented Reality Applications in Medicine: Analysis of the Scientific Literature’, \textit{J. Med. Internet Res.}, vol. 23, no. 2, p. e25499, Feb. 2021.
\bibitem{b7} S. Condino \textit{et al.}, ‘How to Build a Patient-Specific Hybrid Simulator for Orthopaedic Open Surgery: Benefits and Limits of Mixed-Reality Using the Microsoft HoloLens’, \textit{J. Healthc. Eng.}, vol. 2018, pp. 1–12, Nov. 2018.
\bibitem{b8} O. M. Tepper \textit{et al.}, ‘Mixed Reality with HoloLens: Where Virtual Reality Meets Augmented Reality in the Operating Room’, \textit{Plast. Reconstr. Surg.}, vol. 140, no. 5, pp. 1066–1070, Nov. 2017.
\bibitem{b9} P. Milgram and F. Kishino, ‘A taxonomy of mixed reality visual displays’, \textit{IEICE Trans. Info. Syst.}, vol. 77, no. 12, pp. 1321–1329, 1994. 
\bibitem{b10} R. Kolecki \textit{et al.}, ‘Assessment of the utility of Mixed Reality in medical education’, \textit{Transl. Res. Anat.}, vol. 28, p. 100214, Sep. 2022.
\bibitem{b11} C. Lugaresi \textit{et al.}, ‘Mediapipe: A framework for building perception pipelines’, \textit{arXiv preprint} arXiv:1906.08172, 2019.
\bibitem{b12} L. Coyne, J. K. Takemoto, B. L. Parmentier, T. Merritt, and R. A. Sharpton, ‘Exploring virtual reality as a platform for distance team-based learning’, \textit{Curr. Pharm. Teach. Learn.}, vol. 10, no. 10, pp. 1384–1390, 2018. 
\bibitem{b13} M. Dávideková, M. Mjartan, and M. Greguš, ‘Utilization of Virtual Reality in Education of Employees in Slovakia’, \textit{Procedia Comput. Sci.}, vol. 113, pp. 253–260, 2017. 
\bibitem{b14} S. Jayaram, H. I. Connacher, and K. W. Lyons, ‘Virtual assembly using virtual reality techniques’, \textit{Comput.-Aided Des.}, vol. 29, no. 8, pp. 575–584, 1997. 
\bibitem{b15} P. Piromchai, A. Avery, M. Laopaiboon, G. Kennedy, and S. O’Leary, ‘Virtual reality training for improving the skills needed for performing surgery of the ear, nose or throat’, \textit{Cochrane Database Syst. Rev.}, no. 9, 2015. 
\bibitem{b16} K. E. Laver, D. Schoene, M. Crotty, S. George, N. A. Lannin, and C. Sherrington, ‘Telerehabilitation services for stroke’, \textit{Cochrane Database Syst. Rev.}, no. 12, p. CD010255, Dec. 2013.
\bibitem{b17} R. S. Haluck and T. M. Krummel, ‘Computers and virtual reality for surgical education in the 21st century’, \textit{Arch. Surg.}, vol. 135, no. 7, pp. 786–792, 2000. 
\bibitem{b18} A. Blaga and L. Tamas, ‘Augmented reality for digital manufacturing’, in \textit{Proc. 26th Mediterr. Conf. Control Autom. (MED)}, IEEE, 2018, pp. 173–178. 
\bibitem{b19} G. Coelho \textit{et al.}, ‘Augmented reality and physical hybrid model simulation for preoperative planning of metopic craniosynostosis surgery’, \textit{Neurosurg. Focus}, vol. 48, no. 3, p. E19, 2020. 

\bibitem{b20} V. Ferrari \textit{et al.}, ‘Augmented reality visualization of deformable tubular structures for surgical simulation’, \textit{Int. J. Med. Robot.}, vol. 12, no. 2, pp. 231–240, 2016.
\bibitem{b21} M. Ortiz-Catalan \textit{et al.}, ‘Phantom motor execution facilitated by machine learning and augmented reality as treatment for phantom limb pain: a single group, clinical trial in patients with chronic intractable phantom limb pain’, \textit{Lancet}, vol. 388, no. 10062, pp. 2885–2894, Dec. 2016.
\bibitem{b22} W. Barfield, Fundamentals of wearable computers and augmented reality. CRC press, 2015. 
\bibitem{b23} M. Parveau and M. Adda, ‘3iVClass: a new classification method for virtual, augmented and mixed realities’, \textit{Procedia Comput. Sci.}, vol. 141, pp. 263–270, 2018. 
\bibitem{b25} J. L. McJunkin \textit{et al.}, ‘Development of a Mixed Reality Platform for Lateral Skull Base Anatomy’, \textit{Otol. Neurotol.}, vol. 39, no. 10, pp. e1137–e1142, Dec. 2018.
\bibitem{b26} R. Wierzbicki \textit{et al.}, ‘3D mixed-reality visualization of medical imaging data as a supporting tool for innovative, minimally invasive surgery for gastrointestinal tumors and systemic treatment as a new path in personalized treatment of advanced cancer diseases’, \textit{J. Cancer Res. Clin. Oncol.}, vol. 148, no. 1, pp. 237–243, 2022. 
\bibitem{b27} C. Scherl \textit{et al.}, ‘Augmented Reality with HoloLens in Parotid Tumor Surgery: A Prospective Feasibility Study’, \textit{ORL}, vol. 83, no. 6, pp. 439–448, Mar. 2021.
\bibitem{b28} L.-Y. Zhu \textit{et al.}, ‘Application value of mixed reality in hepatectomy for hepatocellular carcinoma’, \textit{World J. Gastrointest. Surg.}, vol. 14, no. 1, p. 36, 2022. 
\bibitem{b29} A. Yamazaki \textit{et al.}, ‘Patient-specific virtual and mixed reality for immersive, experiential anatomy education and for surgical planning in temporal bone surgery’, \textit{Auris. Nasus. Larynx}, vol. 48, no. 6, pp. 1081–1091, 2021.
\bibitem{b30} Lymphatics of head and neck - Download Free 3D model by University of Dundee, School of Dentistry (@DundeeDental), (Nov. 08, 2016). Accessed: Apr. 18, 2025. [Online Video]. Available: https://sketchfab.com/models/5abcaca1574248c0860ae1d81b06998d/
\bibitem{b31} A. Zhang, Z. Min, Z. Zhang, Y. Wang, and M. Q.-H. Meng, “A novel augmented reality assisted orthopedic surgical robotic system with bidirectional surface registration algorithms,” \textit{IEEE Transactions on Medical Robotics and Bionics}, vol. 6, no. 4, pp. 1555–1566, Nov. 2024. doi:10.1109/tmrb.2024.3472844 
\bibitem{b32} J. Fotouhi et al., “Reflective-AR display: An interaction methodology for virtual-to-real alignment in Medical Robotics,” \textit{IEEE Robotics and Automation Letters}, vol. 5, no. 2, pp. 2722–2729, Apr. 2020. doi:10.1109/lra.2020.2972831 
 \end{thebibliography}



\clearpage

\end{document}